\documentclass[useAMS,usenatbib,times,mathptm]{mn2e}
\usepackage{psfig}
\def\mch{M$\rm^{c}$Hardy\,}
\def\etal{et al.~\rm}

\def\mch{M$\rm^{c}$Hardy\,}
\def\etal{et al.~}

\def\mcg6{MCG--6-30-15}

\def\msun{$M_{\odot}$}
\def\ltsim{\mathrel{\hbox{\rlap{\hbox{\lower4pt\hbox{$\sim$}}}\hbox{$<$}}}}
\def\gtsim{\mathrel{\hbox{\rlap{\hbox{\lower4pt\hbox{$\sim$}}}\hbox{$>$}}}}

\newcommand{\aap}{A\&A}
\newcommand{\aapr}{A\&ARv}

\newcommand{\aj}{AJ}
\newcommand{\apj}{ApJ}
\newcommand{\apjl}{ApJL}
\newcommand{\apjs}{ApJS}

\newcommand{\mnras}{MNRAS}
\newcommand{\nat}{Nat}

\begin{document}

\title[Simultaneous X-ray and infrared variability in the
quasar 3C273 II]
{Simultaneous X-ray and infrared variability in the
quasar 3C273 II : Confirmation of the correlation and X-ray lag.}

\author[I.M. M$\rm^{c}$Hardy \etal]
{Ian M$\rm^{c}$Hardy$^{1}$, Anthony
Lawson$^{1}$\thanks{Currently at QinetiQ, St Andrews Road, Malvern, WR14 3PS }, Andrew
Newsam$^{1,2}$, Alan P. Marscher$^{3}$, \and Andrei S. Sokolov$^{3}$, C. Megan
Urry$^{4}$, Ann E. Wehrle$^{5}$\\
$^{1}$ School of Physics and Astronomy, University of Southampton,
SO17 1BJ.\\
$^{2}$ Liverpool John Moores University, Astrophysics Research Institute,
  Birkenhead CH41 1LD \\
$^{3}$ Institute for Astrophysical Research, 725 Commonwealth Avenue,
Boston University, Boston, MA, 02215, USA.\\
$^{4}$
Department of Physics, Yale University, P.O. Box 208121, New Haven
CT 06520-8121, USA\\
$^{5}$ IPAC, Jet Propoulsion Laboratory and California Institute of
Technology, Pasadena, CA 91125, USA.}

\maketitle
\begin{abstract}

The X-ray emission from quasars such as 3C273 is generally agreed to
arise from Compton scattering of low energy seed photons by
relativistic electrons in a relativistic jet oriented close to the
line of sight. However there are a number of possible models for the
origin of the seed photons. In Paper I (\mch \etal 1999) we showed
that the X-ray and IR variability from 3C273 was highly correlated in 1997,
with the IR flux leading the X-rays by $\sim0.75 \pm 0.25$ days. The
strong correlation, and lag, supports the Synchrotron Self-Compton
(SSC) model, where the seed photons are synchroton photons from the
jet itself.

The previous correlation was based on one moderately well sampled
flare and another poorly sampled flare, so the possibility of chance
correlated variability exists.  Here we report on further X-ray and IR
observations of 3C273 which confirm the behaviour seen in Paper I.
During a 2 week period of observations we see a flare of amplitude
$\sim 25\%$, lasting for $\sim5$ days, showing a high correlation
between IR and X-ray variations, with the X-rays lagging by $\sim1.45
\pm 0.15$ days.  These observations were not scheduled at any special
time, implying that the same mechanism - almost certainly SSC -
dominates the X-ray emission on most occasions and that the structure
of the emission region is similar in most small flares.

\end{abstract}


\section{INTRODUCTION}

The quasar 3C273 is a non-extreme, but very bright, example of the
class of active galactic nuclei known as blazars. Blazars are
characterised by high polarisation \citep{valtaoja90} and violent
variability at optical wavelengths. In the case of 3C273, the blazar
nature is more apparent at near-infrared than at optical wavelengths
\citep{robson93}. Blazars contain a strong, relativistically flowing, nonthermally
emitting jet, which features superluminal apparent motion of
parsec-scale radio components
\citep[e.g.,][and references therein]{jorstad01a, jorstad05, mantovani99}.
It is almost universally accepted that the
radio through optical emission in such quasars is synchrotron emission
from a relativistic jet oriented close to the line of sight.  The
smooth synchrotron spectrum does not generally extend to the X-ray
emission, which is best explained by Compton scattering of seed
photons by the relativistic electrons in the jet.
3C273 has been widely observed at X- and Gamma-ray energies
\citep[see, e.g.,][]{courvoisier98,courvoisier03} but the origin of the
seed photons, particularly for the higher energies, remains one of the
major questions in quasar physics.

As well as emission from the relativistic jet, 3C273 also contains a
Seyfert-like nucleus which contributes unbeamed X-ray emission,
\citep[e.g.,][]{grandi04}, perhaps accounting for $\sim20\%$ of the total X-ray
flux. The accretion disc surrounding the central black
hole produces strong UV and optical emission \citep[e.g., see][]{courvoisier03},
and infrared emission in 3C273 can arise from synchrotron radiation from
the jet, free-free emission from dense clouds \citep{robson86} on
sub-pc scales, or
thermal emission from a hot dusty torus on a $\sim$10pc scale
\citep[see][and references therein]{barvainis87,robson93,sokolov05}.
Thus the
overall model for the continuum emission from 3C273 is complex.
In order to investigate the underlying physics of the quasar, and
particularly the origin of the X-ray emission, we must
therefore adopt a method that can isolate the different components.
Our approach is to study variability of the continuum emission in
different wavebands, looking for correlations that reveal the primary
emission mechanisms and their locations in the quasar (e.g., jet or
Seyfert-like nucleus). In order to search particularly for a
relationship between the Compton-scattered X-ray photons and possible
synchrotron seed photons we \cite[Paper I]{mch99} carried out a
programme of correlated near-infrared (K band, 2.2 $\mu$m)  and
medium energy X-ray (3-20 keV) observations of 3C273 in 1996/7.
These observations
revealed a strong correlation between the emission in the two bands
and showed that the X-rays lag the IR by $0.75\pm0.25$days.

These previous observations, although the first serious IR/X-ray
variability observations of a blazar, were not extensive. One X-ray
flare was reasonably covered by IR observations and a second was
sparsely covered.  However it was not clear whether these observations were
representative of the general IR/X-ray behaviour of 3C273. It is
possible that the IR and
X-ray variations may not have been physically correlated and may have
been merely chance events \citep[e.g. see discussion in][]{mason02}.
In order to clarify whether the previously
observed IR/X-ray correlation was, indeed, representative of the
general behaviour of 3C273, we therefore repeated the experiment in
March 1999, with better temporal coverage.

Here we report the result of the March 1999 observations, during which a flare
was seen first at K-band and then, with a delay of $1.45\pm0.15$ days,
in X-rays. We discuss the combined significance of the 1996/7 and 1999
observations and consider various theoretical
interpretations. We show that the most reasonable explanation of the
X-ray variability is synchrotron self-Compton scattering.  In Section
2 we describe the observations and in Section 3 we determine the lag
between the X-ray and IR bands. In Section 4 we compare our
observations with theoretical models and in Section 5 we summarise our
conclusions.

\section{OBSERVATIONS: Second Campaign}

\begin{figure*}
\psfig{figure=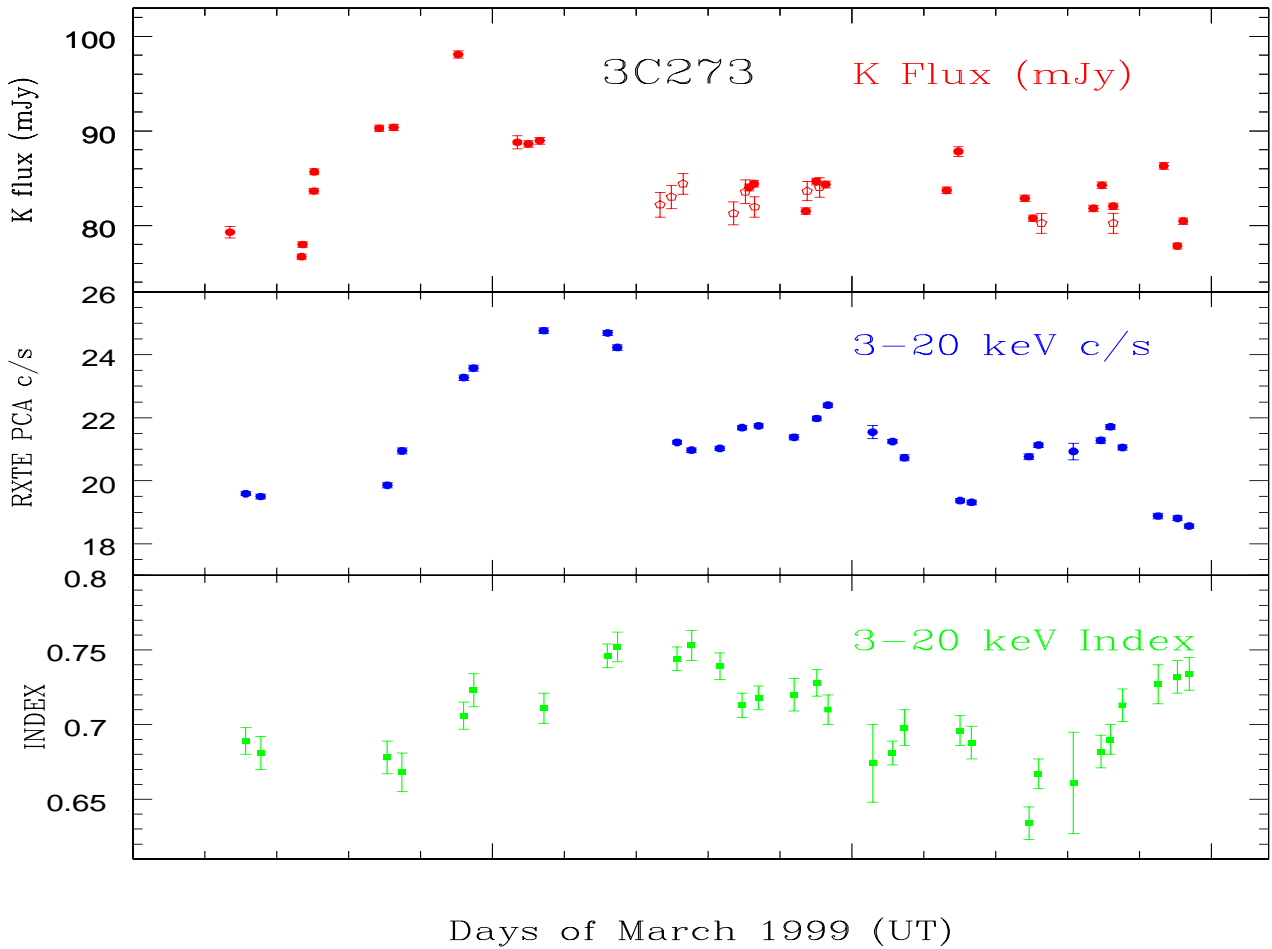,height=13.5cm,width=18cm}
\caption{X-ray and near-infrared K-band lightcurves and X-ray spectral ``energy''
index as a function of time.
The X-ray counts are the total from 2 PCUs of the PCA.
The time covered by each data point is between 4 and 8 ksec.
Where definite variability is seen within an observation,
two data points are given; otherwise only one is given.
The errorbars are plotted, but are generally too small to see.
The K-band data from UKFTI on UKIRT are displayed as filled circles,
the data from IRCAM3 on UKIRT are open pentagons.
Note that the dates are labelled such that 0 March means
00:00 on 1 March 1999.
 }
\label{fig:lcurves99}
\end{figure*}

\subsection{X-ray Observations}

We observed 3C273 from 1 to 15 March 1999 with the Proportional
Counter Array (PCA) on RXTE. The only significant difference from the
1996/7 observations is that only 2 of the Proportional Counter Units
(PCUs) of the PCA were in operation, compared to 3 in the previous
observations, hence the total count rate is down by 33\% relative to Paper I
for the same photon flux. The data were reduced in exactly the same standard way
as is described in Paper I, where the reader can find details of the analysis.

We observed the source for at total of $\sim250$ksec,
with approximately 5 hours per day on target. The observation time per
day was therefore much more than in the
previous observations, although
the 5 hours were not usually spread evenly throughout the day
but typically occurred in 2 or 3 blocks. After rejection of
data that did not satisfy the standard acceptance criteria, a total
of 172~ksec of good-quality data were retained.
X-ray variability on very short timescales ($<1$ksec) is of low
amplitude in 3C273, so these data were split into segments of $\sim5$ksec duration.
In Fig~\ref{fig:lcurves99} we show the resultant 3-20 keV count rate lightcurve.  As
in Paper I, continuous X-ray variability is seen on all timescales
longer than the bin size (5~ksec).  The longer observation times
compared to the 1996/7 observations result in smaller flux errors,
which allows us to detect real variability at the few percent level on
timescales of $\sim$hour. As with Seyfert galaxies, 3C273 shows
continual variations with larger amplitude variability on longer
timescales. There is a lack of flickering on very short timescales ($<$hours)
compared to Seyfert galaxies. This may be explained simply
by a higher black hole mass, and therefore larger characteristic size
scales, than in Seyfert galaxies \citep[e.g.,][]{mch04,mch05}, but we
have not yet completed a proper power-spectral analysis of the X-ray
variability, and so are unable to make a quantitative comparison.

\begin{figure*}
\psfig{figure=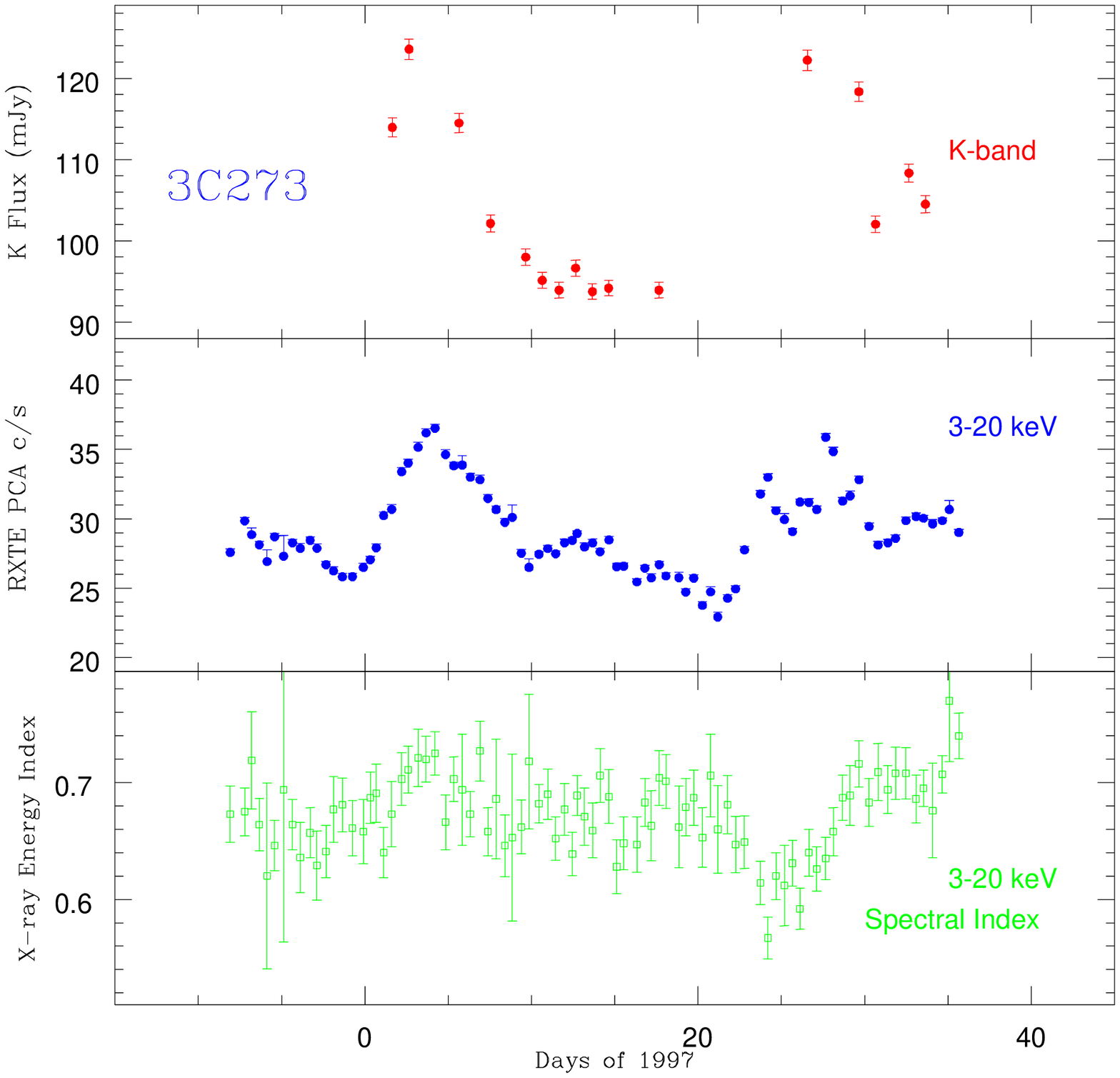,height=13.5cm,width=18cm}
\caption{
X-ray and near-infrared K-band lightcurves and X-ray spectral
``energy'' index as a function of time.  The X-ray counts are
normalised to 2 PCUs, for consistency with Fig~\ref{fig:lcurves99}.
The time covered by each X-ray data point is 1 ksec.  The errorbars
are plotted, but are generally too small to see.  Note that the dates
are labelled such that 0 means 00:00 on 1 January 1997.}
\label{fig:lcurves97}
\end{figure*}

The observations in 1996/7 occurred when 3C273 was a little brighter,
and the flares were of slightly larger amplitude, than in March 1999.
The mean level in
Fig~\ref{fig:lcurves99} is approximately 75\% of the mean level
shown in Paper I (reproduced here in Fig.~\ref{fig:lcurves97}).
The X-ray flare beginning on 3 March 1999 has an
amplitude of $\sim25\%$ of the mean level and is discernible above the
mean level for about 4 days, compared to a flare amplitude of
$\sim40\%$ of the mean for the two flares in
Paper I, each of which was discernible above the mean for about 10 days.

\subsection{Near-Infrared Observations}

K-band observations were made with the UFTI and IRCAM3 imaging cameras
at UKIRT at the start, middle, and end of each night from 3 to 15 March
1999, thereby improving temporal coverage by a factor of 3 compared to
the 1997 observations.  Typical exposures were 3 minutes and were
reduced in the standard manner, as described in Paper I. The resultant
lightcurve is shown in Fig.~\ref{fig:lcurves99}.  IRCAM K-band fluxes are transformed
to the UFTI K-band by assuming a colour for 3C273 of J-K = 2 mag, the
value that we have measured in previous observations.
Uncertainties in the transformation are included in the errors.  Note
that for the very first IR observation the calibration star was 4
magnitudes fainter than for all of the other observations, and so the
errorbars may be underestimated.

Additional K-band observations were made at the 0.75m telescope of the
South African Astronomical Observatory (SAAO) during the UKIRT day in
order to improve temporal coverage still further.  Unfortunately,
despite longer integration times, the resultant errors were factors of
a few larger than the errors on the UKIRT observations, and so we
do not include these data in the present analysis.

The large flare visible at the beginning of the X-ray lightcurve
(Fig~\ref{fig:lcurves99}) is
also clearly seen in the IR lightcurve and
appears to precede the X-ray lightcurve by about a day. After 8 March
the X-ray and IR lightcurves display only 5-10\% variability, which
does not appear to be strongly correlated between the bands, although
we note that the dip in the X-ray lightcurve on 12 March would not
have been sampled by the IR lightcurve had it occurred on 11 March.

\section{X-RAY/INFRARED CROSS-CORRELATION}

\begin{figure}
\psfig{figure=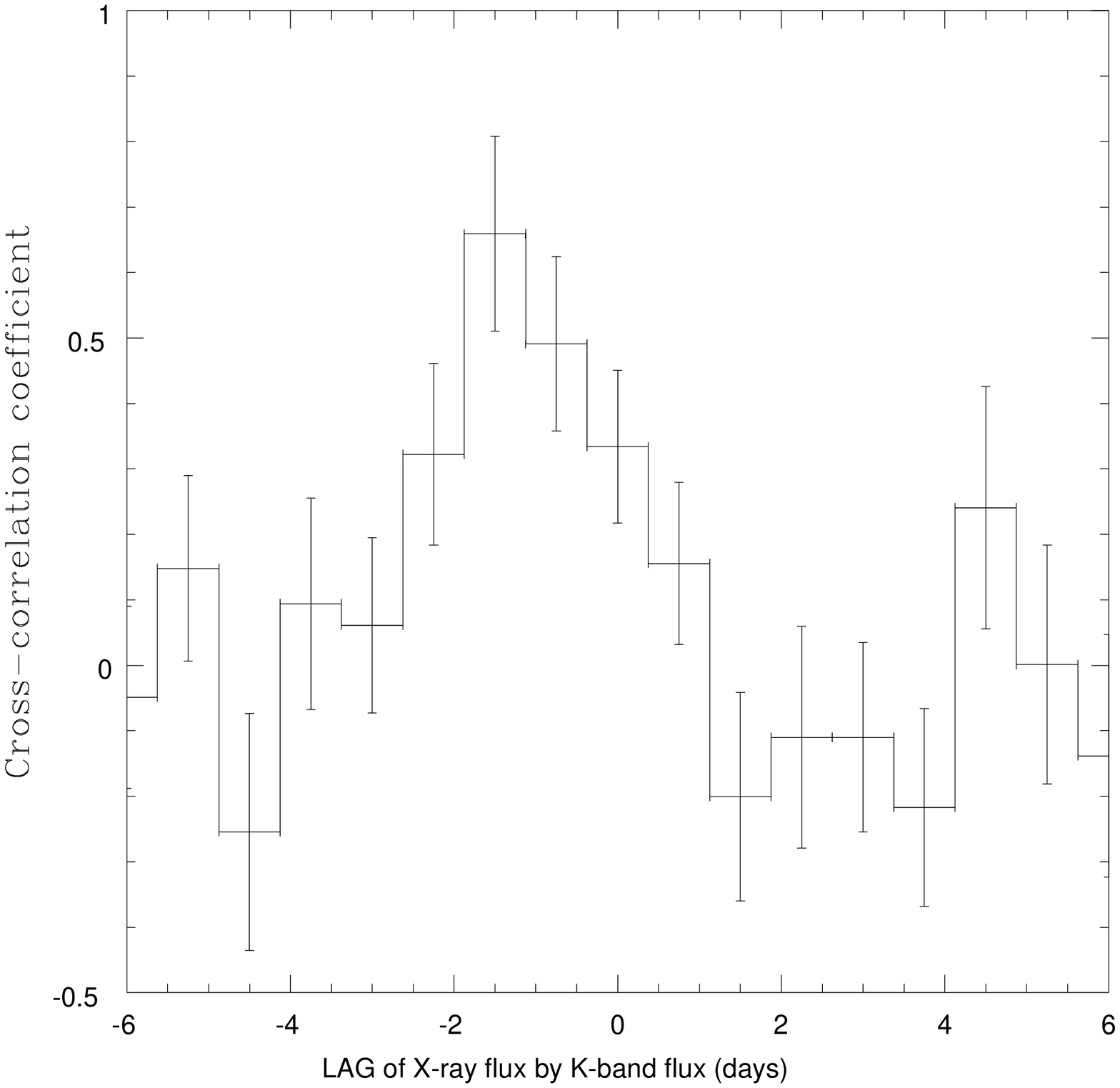,height=8cm,width=8.5cm}
\caption{Cross-correlation of the X-ray and IR fluxes. The bin size
is 0.75 days. }
\label{fig:dcf}
\end{figure}

We follow the same procedure as given in Paper I to determine
the correlation between the X-ray and IR fluxes. We first perform a
discrete cross-correlation \citep{edelson88}, using all of the
observations presented in Fig~\ref{fig:lcurves99}. The result is given
in Fig~\ref{fig:dcf}
which shows that the X-rays lag the IR by approximately 1.5 days.

In order to refine the measurement of the lag we perform the same
fitting operation that we used in Paper I.  Assuming that the X-rays
result from scattering of the IR photons, we can parameterise the
relationship by:

\[ X_{predicted}(t)= A \, (K_{flux}(t-\delta t) - K_{quiescent})^{N}
\, + \, X_{quiescent}. \]
$K_{quiescent}$ is a non-varying K-band component, probably from
dust emission, of amplitude 50 mJy \citep{robson93}.  $K_{flux}(t-\delta t)$
is the total observed K-band flux at time $t-\delta t$ and
$X_{predicted}(t)$ is then the predicted total X-ray flux at time $t$.
$A$ is the constant of proportionality
and $N$ contains information about the emission mechanism. For
example, in the SSC process if the X-rays arise from variations in
the number of high-energy electrons then we expect $N=2$ , but
$N=1$ if the variations result from changes solely in magnetic
field strength.

$X_{quiescent}$ is the part of the X-ray flux that does not come
from the flaring region. This emission could come from the
Seyfert-like nucleus, or from the `quiescent' jet. \cite{grandi04}
show that only $\sim20\%$ of the X-ray emission in 3C273 comes
from the Seyfert-like nucleus. As the black hole in 3C273 is very
large \citep[$6 \times 10^{9}$ \msun,][]{paltani05}, large amplitude variations of the Seyfert
component on $\sim$week timescales will be unimportant 
 \citep[e.g. see][]{mch05} and, by
definition, variations from the `quiescent jet' will be on longer
timescales than those discussed here. Thus, although a slight
variation of $X_{quiescent}$ may occur over the $\sim$weeks
timescale of our observations, it is adequate for the purposes of
examining only the larger, and more pronounced variations, to
treat $X_{quiescent}$ as a constant. We have experimented with
allowing $X_{quiescent}$ to vary with a linear trend, and the best
fit is for a decrease, in the 1996/7 observations, of about 10\%
over the observations. But the derived values of the other
parameters do not change significantly and the improvement in fit
is not sufficient to justify the inclusion of another free
parameter.

\begin{figure}
\psfig{figure=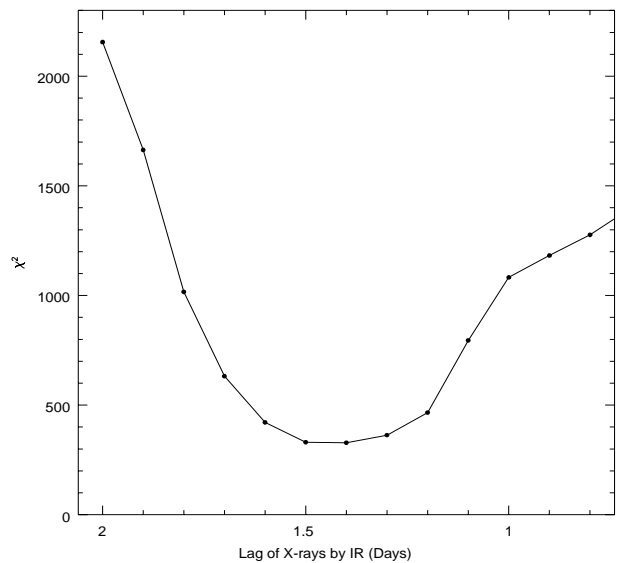,height=8cm,width=8.5cm}
\caption{
Result (in terms of the $\chi^2$ statistic)
of comparing the observed X-ray lightcurve with that predicted
from the infrared variations as a function of lag. For each value of the lag,
all other parameters are allowed to remain free.
}
\label{fig:lagschi}
\end{figure}

The variable $\delta t$ allows for lags between the X-ray and
infrared variations. Neither the previous nor present observation
allows us to tightly constrain $N$, $A$, or $X_{quiescent}$, but
varying $\delta t$, whilst allowing the other variables to remain
free, does affect the fit significantly. As in Paper I we
therefore perform a $\chi^{2}$ fit, comparing the predicted X-ray
flux with the observed flux, using a standard Levenburg-Marquardt
minimisation routine, for various values of $\delta t$. We apply
simple linear interpolation to estimate the X-ray flux at the
exact (shifted) time of the IR observations. \footnote[1]{Note, in
Paper I we followed exactly the same procedure but the paper
incorrectly states that we interpolated the IR flux to the exact
(shifted) time of the X-ray observations. In any case, the results
do not depend significantly on whether one interpolates the IR or
the X-ray data.} The results are shown in
Figure~\ref{fig:lagschi}.  In this fit we use only the data prior
to 8 March, covering the large flare.  If we use all of the data
the fit becomes worse as it becomes dominated by the relatively
quiescent levels after March 8, although it still indicates that
the IR leads the X-rays.

The minimum $\chi^{2}$ value occurs when the IR leads the X-rays by
1.45 days. The high absolute values of $\chi^{2}$ reflect the fact
that the errors of both the X-ray and IR fluxes are low and that,
although the flare is broadly the same shape in both bands, the
detailed behaviour is not exactly the same. In other words the model,
although reasonable to zeroth order, does not provide a detailed
description of the observed
behaviour. Therefore, we cannot obtain an error on the lag by simply
measuring the difference in lag at some value of $\delta \chi^{2}$
away from the minimum which, for a good model fit, might give us a
90\% confidence region (see \cite{press92}, Section 15.6, for
details).  We have therefore carried out a Monte Carlo simulation,
scattering all the data points by their errors and refitting. After
10,000 such simulations, we derive an error of 0.05 days for the lag.

\begin{figure}
\psfig{figure=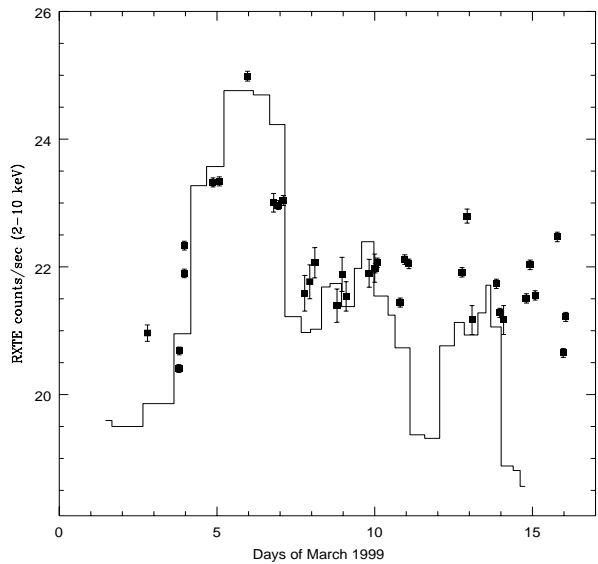,height=8cm,width=8.5cm}
\caption{Observed X-ray lightcurve (histogram) and the best-fit
predicted X-ray flux (filled squares) based on the parameters derived
from fitting the infrared observations to the initial flare (20 IR
data points) only. The best-fit parameters are $A=0.24$, $N=0.98$ and
$X_{quiescent}=14.5$.  Following from Fig.~\ref{fig:lagschi} the
lead of the IR over the observed X-rays is fixed at 1.45 days, the
best-fit value for the first flare.  The observed X-ray errorbars (see
Fig.~\ref{fig:lcurves99}) are not repeated here to avoid cluttering
the diagram.  }
\label{fig:xpred}
\end{figure}

Using the parameters and lag defined by our best fit, we predict the
X-ray flux and show it, superposed on the observed flux, in
Fig.~\ref{fig:xpred}. As in Paper I we see that the fit to the flare
at the start of the observational period is good but once the flare
has faded the correspondence becomes looser, presumably as additional
minor components, e.g., the Seyfert-like nucleus, also contribute
noticeably to the flux \citep[see][]{grandi04}. Our sole interest here is
to analyse the most pronounced flares, hence we will ignore the post-flare
behaviour, our data for which are too limited to interpret.

\section{COMPARISON WITH THEORETICAL MODELS}

In both 1997 and 1999 we have observed reasonably well resolved flares
in 3C273 where the X-ray emission lags the IR emission by about a day
and, on both occasions, the X-ray spectrum of the flaring emission was
softer than the quiescent emission. Staring around day 20 of 1997 we
also note a second, more irregular period of enhanced X-ray emission
where the lag is probably shorter and where the X-ray spectrum becomes
harder during the enhanced emission. In this section we consider the
implications of these observations for theoretical models of its X-ray
emission and for the specific structure
of the jet and X-ray emission region in 3C273.

\subsection{Summary of X-ray Emission Mechanisms}
A perturbation, e.g., a shock or other compression, in the jet leading
to higher electron density and enhanced synchrotron emission would
lead to greater scattering of the ambient nuclear
UV/optical/IR photon field, resulting in an increase in X-ray and
Gamma-ray flux
\citep[the External Compton, or EC, model, ][]{dermer93,sikora94,blaz00}.
However, the synchrotron photons
must also be scattered by their parent electrons
\citep[the Synchrotron Self-Compton, or SSC, model, ][]{Jones74a},
producing an elevated level of high-energy emission. One of our main
aims here is to distinguish which mechanism is more important for the
nonthermal X-ray emission in 3C273.

A previously-popular variant of
the SSC model was the Mirror Compton (MC) model \citep{ghisellini96}.
Here the synchrotron photons from the jet are reflected back to the
jet from a neighbouring cloud with the scattered emission lagging the
synchrotron emission by approximately the light travel distance
to the cloud. However detailed calculation shows that any reflecting
cloud would need to lie essentially in the jet to produce a significant
scattered high-energy flare \citep{boettcher98,bednarek98}.
Thus the MC model is unlikely to be the cause
of the majority of X-ray flares.

\subsection{X-ray/IR lags as Emission Diagnostics}
There are a number of variables, e.g. the geometry of the
emitting region and the angle of the jet to the line of sight, which
can affect the decay timescales, amplitudes and lag between the X-ray
and IR emission. We refer to \cite{sokolov04} for a general discussion of
the modelling of nonthermal flares similar to those shown here. Here, at the
risk of some oversimplification, we concentrate on the two general
points of most relevance to observations.

Firstly, in the EC process, the electrons responsible for
synchrotron IR flares have high energies ($\gamma \sim 1000$) and so
decay quickly, whereas the electrons responsible for scattering
IR/optical/UV photons to X-ray energies are of low
energy ($\gamma \sim 10$) and so have much longer decay timescales.
Thus the X-ray decay timescale is much
longer than the IR decay timescale and so we do not expect flares to
look similar in both bands. However in the SSC process, higher energy
electrons contribute strongly to the X-ray emission (see, e.g., Paper I) and so we
expect the X-ray and IR flares to have moderately similar shapes and
be better correlated than in the EC case. The similarity of our
observed X-ray and IR flares therefore favours the SSC model.

Secondly, in the EC process, the synchrotron and X-ray variations
should rise almost simultaneously as the ambient UV/optical/IR field
is all-pervasive and is immediately available for scattering to X-ray
energies. However in the SSC process there are two reasons why
high-energy flares might lag IR or optical synchrotron
flares.  If the jet points almost along the line of sight,
then the seed photon field does not rise instantaneously at the
location of any given electron, but reaches a peak after
approximately half of the light travel time across the emission
region. The X-ray emission should therefore lag the
synchrotron emission by about the same time. The more general reason for a
lag is frequency stratification within the source
\citep{marscher85}. The SSC flux at any given X-ray energy is produced by a
combination of electron energies and seed-photon frequencies,
including relatively low-energy electrons scattering high frequency
photons to high-energy electrons scattering low frequency photons (see
Fig.~3 in Paper I). If the higher-energy electrons and high frequency
synchrotron photons are confined to a small volume, e.g. close to a
shock front, the X-ray flare, which comes from a more extended region,
will have a time-delayed maximum (and may also be more prolonged than
the highest frequencies in the synchrotron flare). The delay of the
X-ray relative to the IR therefore again favours the SSC over the EC
model for the origin of the X-ray emission. 

The 2-10 keV emission arises from self-Compton scattering of a large
range of seed photon energies, from radio through to IR, by a large
range of electron energies (see Paper 1). However high energy
Gamma-ray emission is produced by scattering, by the highest energy
electrons, of the higher energy seed photons (optical/UV). The
Gamma-ray emission therefore need not necessarily be dominated by SSC
emission but could arise, at the higher (GeV) energies from scattering
of ambient optical/UV photons from the accretion disc and broad line
region
\citep{kubo98,dermer93,sikora94} or, at MeV energies, even by scattering of
IR emission from the torus \cite{blaz00}.

\subsection{The Basic Jet Structure} The short-timescale near-infrared flares
are almost certainly enhancements of the synchrotron emission from the
jet.  The timescale of variability of $\sim2$ days implies that the
flaring region has a maximum size of $\sim2\delta \sim 20$lt-days
$\sim 0.02$pc, where the value of the relativistic Doppler factor,
$\delta\approx9$ (Lorentz factor 11, angle to the line of sight
$6^{\circ}$)
in the jet in early 1999 was derived by
\citet{jorstad05} from apparent superluminal motion and the decay time
of the flux of moving radio knots. These authors also determined the opening
half-angle of the jet to be $1.4^\circ \pm0.3^\circ$. If the jet has a
conical geometry, the distance from the apex corresponding to 0.02 pc is
$\sim0.5$ pc, or greater if the flaring region does not span the entire
width of the jet.

\subsection{X-ray spectral variability: the emission region}
Precise interpretation of X-ray spectral variability is difficult as
we are almost certainly seeing the superposition of a number of
different events, so we restrict ourselves to commenting only on the
larger events.

In the March 1999 flare (Fig.~\ref{fig:lcurves99}) and in the first of
the January 1997 flares (Fig.~\ref{fig:lcurves97}) we note broadly
similar spectral variability. In each case the spectrum of the flare
is softer than that of the quiescent state, which is most simply
explained if the injected electron energy distribution is steeper than
that of the quiescent jet. Such an interpretation is consistent with
the observation that the X-ray spectral index returned to the previous
value as the flare died out. We do note, in the better sampled March
1999 flare, that the peak in the spectral index lags that in X-ray
flux by about a day or so, probably indicating ageing of the flare
electron distribution caused by radiative losses.

However in the second 1997 flare we note the reverse behaviour as the
spectrum initially hardens with increasing flux before softening back
towards the quiescent value.  This spectral difference may be related
to temporal differences. Unlike the smooth first flare, the second
flare appears to be made up of three or four smaller flares,
superposed, and the rise of the second flare is more abrupt than that
of the first flare. Also, although not as well constrained by
observation, we noted in Paper I that the $\sim1$day lag of the IR by
the X-rays during the first flare did not easily apply to the second
flare, and that a zero delay was a better fit. A possible explanation,
therefore, is that in the second 1997 flare we see not just one
instance of excitation (e.g., a shock), but a succession of small
regions of particle acceleration, perhaps with one shock compounding
the energization of the previous one \citep[e.g., as in a colliding
shock model;][]{spada01}. Thus the dominant emission region remains
small, consistent with the very small IR-X-ray lag, and particles are
continually reaccelerated, a situation that favours the acceleration
of electrons to high rather than low eneries. In this scenario, the
X-ray spectral variability reflects the multiplicity of episodes of
re-energisation of the jet plasma. The observed spectral differences
in the flares shown here may therefore simply be reflecting the
spectrum of turbulence in the underlying jet.

\section{CONCLUSIONS}

Our present observations repeat the general pattern of X-ray and IR
variability described in Paper I.
X-ray variability in 3C273 is seen on all timescales which are
currently available to observation, i.e., $>$hours.  We again
demonstrate that there is a strong correlation between variations in
the X-ray and IR bands for variations of amplitude $>$20\%,
confirming that the observations described in Paper I were not the
result of chance coincidence. For lower amplitude
variations the correlation is less clear, probably indicating that
multiple emission regions and processes are involved
\citep[see, e.g.,][]{grandi04,courvoisier98}.

The observations presented here and in Paper I were not scheduled at any
particularly special time, or flux level, and so confirm that the behaviour presented
here, and in Paper I, is typical of 3C273. In other words the same mechanism
dominates the X-ray emission of 3C273 during small flares, and the structure of
the emission region during those flares must be broadly the same.
Slight differences in jet properties, e.g., location and size of the flaring
region, shock strength, bulk Lorentz factor, and direction of the jet,
can easily alter the flare amplitude and lag timescale. We note that the
latter changed only from $0.75\pm0.25$ to $1.45\pm0.15$ days, a quite modest
difference given the number of variables involved.

We note that the profile of temporal variability is linked to how the
X-ray spectrum varies and provides a good indicator of the structure
of the underlying emission region.

The fact that there is again a strong correlation between the X-ray
and IR variations, with flares of broadly similar shapes with the X-ray
maximum lagging the IR peak,
strongly supports the SSC, rather than EC, model for the production of
the X-rays in 3C273.  As in Paper I, our simplistic fitting procedure
does not constrain significantly the parameters $A$, $N$ or
$X_{quiescent}$ although, as in Paper I, values of $N$ nearer to 1
than 2 are preferred indicating, in the SSC model, that variations in
magnetic field strength as well as electron density, are important in the
X-ray variations.\\

{\bf Acknowledgments} We are very pleased to thank the management and
operational staff of both RXTE and UKIRT for their cooperation in
scheduling and carrying out these observations.  IM$\rm ^{c}$H thanks
PPARC for support by grant PP/D001013/1 and by the award of a Senior Fellowship.
The Boston University effort was supported in part by the National
Science Foundation through grants AST-0098579 and AST-0406865 and by NASA through
grants NAG5-11811 and NAG5-13074.

\bibliographystyle{mn2e}

\end{document}